# State-of-the-art in speaker recognition[1]

Marcos Faundez-Zanuy, (*) Enric Monte-Moreno
Escola Universitaria Politècnica de Mataró, (*) TALP research center (BERCELONA)
Avda. Puig i Cadafalch 101-111
08303 MATARO (BARCELONA) SPAIN
E-mail: faundez@eupmt.es, enric@gps.tsc.upc.es  http:www.eupmt.es/veu

**ABSTRACT**
Recent advances in speech technologies have produced new tools that can be used to improve the performance and flexibility of speaker recognition While there are few degrees of freedom or alternative methods when using fingerprint or iris identification techniques, speech offers much more flexibility and different levels for performing recognition: the system can force the user to speak in a particular manner, different for each attempt to enter. Also with voice input the system has other degrees of freedom, such as the use of knowledge/codes that only the user knows, or dialectical/semantical traits that are difficult to forge.
This paper offers and overview of the state of the art in speaker recognition, with special emphasis on the pros and contras, and the current research lines. The current research lines include improved classification systems, and the use of high level information by means of probabilistic grammars. In conclusion, speaker recognition is far away from being a technology where all the possibilities have already been explored.

## INTRODUCTION

Biometric recognition offers a promising approach for security applications, with some advantages over the classical methods, which depend on something you have (key, card, etc.), or something you know (password, PIN, etc.). However, there is a main drawback, because it cannot be replaced after being compromised by a third party. Probably these drawbacks have slowed down the spread of use of biometric recognition [1-2]. For those applications with a human supervisor (such as border entrance control), this can be a minor problem, because the operator can check if the presented biometric trait is original or fake. However, for remote applications such as internet, some kind of liveliness detection and anti-replay attack mechanisms should be provided. Fortunately, speech offers a richer and wider range of possibilities when compared with other biometric traits, such as fingerprint, iris, hand geometry, face, etc. This is because it can be seen as a mixture of physical and learned traits. We can consider physical traits those which are inherent to people (iris, face, etc.), while learned traits are those related to skill acquired along the life and the environment (signature, gait, etc.). For instance, your signature is different if you have born in a western or an Asiatic country, and your speech accent is different if you have grown up in Edinburgh or in Seattle, and although you might speak the same language, probably the prosody or the vocabulary might be different (i.e. the relative frequency of the use of common words might vary depending on the geographical or educational background).

## SPEECH PROCESSING TECHNIQUES

Speech processing techniques relies on speech signals usually acquired by a microphone and introduced in a computer using a digitalization procedure. It can be used to extract the following information from the speaker:
- Speech detection: is there someone speaking? (speech activity detection)
- Sex identification: which is his/her gender? (Male or female).
- Language recognition: which language is being spoken? (English, Spanish, etc.).
- Speech recognition: which words are pronounced? (speech to text transcription)
- Speaker recognition: which is the speaker's name? (John, Lisa, etc,)

Most of the efforts of the speech processing community have been devoted to the last two topics. In this paper we will focus on the latest one, and the speech related aspects relevant to biometric applications.

## SPEAKER RECOGNITION

---

[1] This work has been supported by FEDER and MEC, TIC-2003-08382-C05-02

Speaker recognition can be performed in two different modes:

**Speaker identification:** In this approach no identity is claimed from the speaker. The automatic system must determine who is talking. If the speaker belongs to a predefined set of known speakers it is referred as closed-set speaker identification. However, for sure the set of speakers known (learnt) by the system is much smaller than the potential number of users than can attempt to enter. The more general situation where the system has to manage with speakers that perhaps are not modeled inside the database is referred as open-set speaker identification. Adding a "none-of-the-above" option to closed-set identification gives open-set identification. The system performance can be evaluated using an identification rate.

**Speaker verification:** In this approach the goal of the system is to determine whether the person is who he/she claims to be. This implies that the user must provide an identity and the system just accepts or rejects the users according to a successful or not verification. Sometimes this operation mode is named authentication or detection. The system performance can be evaluated using the False Acceptance Rate (FAR, those situation where an impostor is accepted) and the False Rejection Rate (FRR, those situation where a speaker is incorrectly rejected), also known in detection theory as False Alarm and Miss respectively. This framework gives us the possibility of distinguishing between the discriminability of the system and the decision bias. The discirminability is inherent to the used classification system and the discrimination bias is related to the preferences/necessities of the user in relation to the relative importance of each of the two possible mistakes (misses vs. false alarms) that can be done in speaker identification. This trade-off between both errors has to be established usually by adjusting a decision threshold. The performance can be plotted in a ROC (Receiver Operator Characteristic) or in a DET (Detection error trade-off) plot [3]. DET curve gives uniform treatment to both types of error, and uses a scale for both axes which spreads out the plot and better distinguishes different well performing systems and usually produces plots that are close to linear. Note also that the ROC curve has symmetry with respect to the DET, i.e. plots the hit rate instead of the miss probability, and uses a logarithmic scale that expands the extreme parts of the curve, which are the parts that give the most information about the performance of the system. For this reason the speech community prefers DET instead of ROC plots. Figure 1 shows an example of DET of plot, and figure 2 shows a classical ROC plot.

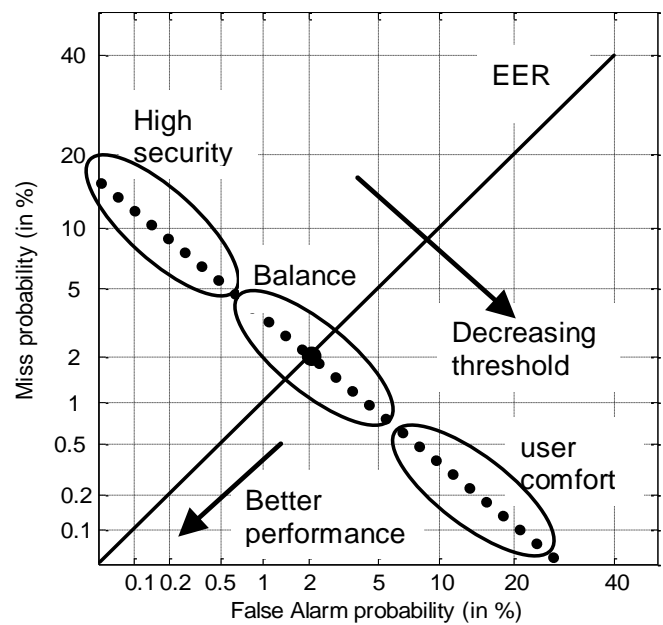

**Figure 1**. Example of a DET plot for a speaker verification system (dotted line). The Equal Error Rate (EER) line shows the situation where False Alarm equals Miss Probability (balanced performance). Of course one of both errors rates can be more important (high security application versus those where we do not want to annoy the user with a high rejection/ miss rate). If the system curve is moved towards the origin, smaller error rates are achieved (better performance). If the decision threshold is reduced, we get higher False Acceptance/Alarm rates.

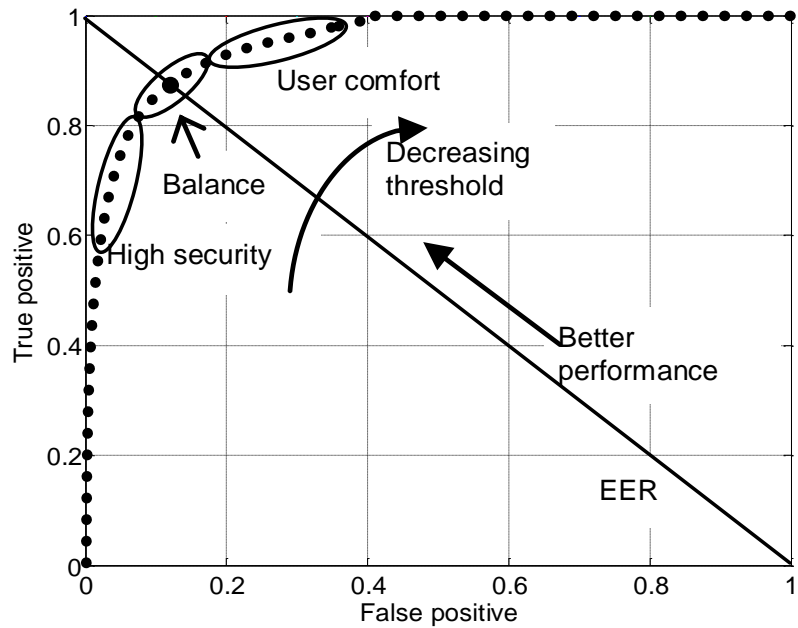

**Figure 2**. Example of a ROC plot for a speaker verification system (dotted line). The Equal Error Rate (EER) line shows the situation where False Alarm equals Miss Probability (balanced performance). Of course one of both errors rates can be more important (high security application versus those where we do not want to annoy the user with a high rejection/ miss rate). If the system curve is moved towards the upper left zone, smaller error rates are achieved (better performance). If the decision threshold is reduced, we get higher False Acceptance/Alarm rates. It is interesting to observe that comparing figures 1 and 2 we get True positive = (1 – Miss probability) and False positive = False Alarm.

In both cases (identification and verification), speaker recognition techniques can be split into two main modalities:

**Text independent:** This is the general case, where the system does not know the text spoken by person. This operation mode is mandatory for those applications where the user does not know that he is being evaluated for recognition purposes, such as in forensic applications, or to simplify the use of a service where the identity is inferred in order to improve the human/machine dialog, as is done in certain banking services. This allows more flexibility, but it also increases the difficulty of the problem. If necessary, speech recognition can provide knowledge of spoken text. In this mode one can use indirectly the typical word co-occurrence of the speaker, and therefore also characterize the speaker by a probabilistic grammar. This co-occurrence model is known as n-grams, and gives the probability that a given set of n words are uttered consecutively by the speaker. This can distinguish between different cultural/regional/gender backgrounds, and therefore complement the speech information, even if the speaker speaks freely. This modality is also interesting in the case of speaker segmentation, when there are several speakers present and there is an interest of segmenting the signal depending on the active speaker.

**Text dependent:** This operation mode implies that the system knows the text spoken by person. It can be a predefined text or a prompted text. In general, the knowledge of the spoken text let to improve the system performance with respect to previous category. This mode is used for those applications with strong control over user input, or in applications where a dialog unit can guide the user.

One of the critical facts for speaker recognition is the presence of channel variability from training to testing. That is, different signal to noise ratio, kind of microphone, evolution with time, etc. For human beings this is not a serious problem, because of the use of different levels of information. However, this affects automatic systems in a significant manner. Fortunately higher-level cues are not as affected by noise or channel mismatch. Some examples of high-level information in speech signals are speaking and pause rate, pitch and timing patterns, idiosyncratic word/phrase usage, idiosyncratic pronunciations, etc.

Looking at the first historical speaker recognition systems we realize that they have been mainly based on physical traits extracted from spectral characteristics of the speech signals. So far, features derived from spectrum of speech have proven to be the most effective in automatic systems, because the spectrum reflects the geometry of the system that generates the signal. Therefore the variability in the dimensions of the vocal tract is reflected in the variability of spectra between speakers [4]. However, there is a large amount of possibilities [5]. Figure 3 summarizes different levels of information suitable for speaker recognition, being the top part related to learned traits and the bottom one to physical traits. Obviously we are not bound to use only one of these levels, and we can use some kind of data fusion [6] in order to obtain a more reliable recognizer.

Learned traits, such as semantics, diction, pronunciation, idiosyncrasy, etc. (related to socio-economic status, education, place of birth, etc.) is more difficult to automatically extract. However, they offer a great potential. For sure, some times when we try to imitate the voice of another person we use this kind of information. Thus, it is really characteristic of each person. Nevertheless, the applicability of these high-level recognition systems is limited by the large training data requirements needed to build robust and stable speaker models. However a simple statistical tool, such as the n-gram can capture easily some of these high level features. For instance in the case of the prosody one could classify a certain number of recurrent pitch patterns, and compute the co-occurrence probability of these patterns for each speaker. This might reflect dialectical and cultural backgrounds of the speaker. From a syntactical point of view this same tool could be used for modeling the different co-occurrence of words for a given speaker.

The interest of making a fusion [6] of both learned and physical traits is that the system is more robust (i.e, increases the separability between speakers), and at the same time is more flexible, because does not force an artificial situation on the speaker. On the other hand, the use of learned traits such as semantics, or prosody introduces a delay on the decision because of the necessity of obtaining enough speech signal for computing the statistics associated to the histograms.

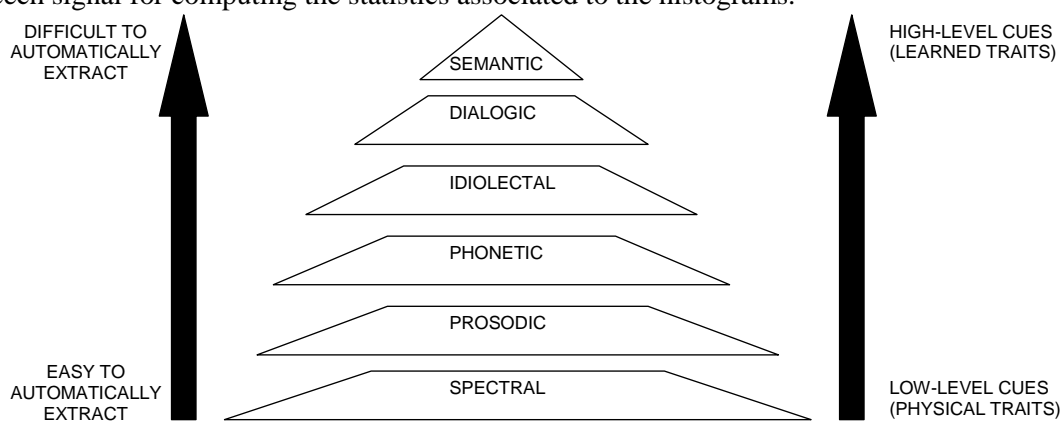

**Figure 3.** Levels of information for speaker recognition

Different levels of extracted information from the speech signal can be used for speaker recognition. Mainly they are:

**Spectral:** The anatomical structure of the vocal apparatus is easy-to-extract in an automatic fashion. In fact, different speakers will have different spectra (location and magnitude of peaks) for similar sounds. The state-of-the-art speaker recognition algorithms are based on statistical models of short-term acoustic measurements provided by a feature extractor. The most popular model is the Gaussian Mixture Model (GMM) [7], and the use of Support Vector Machines [8]. Feature extraction is usually computed by temporal methods like the Linear Predictive Coding (LPC) or frequencial methods like the Mel Frequency Cepstral Coding (MFCC) or both methods like Perceptual Linear Coding (PLP). A nice property of the spectral methods is that logarithmic scales (either amplitude or frequency), which mimic the functional properties of the human ear, improve the recognition rates. This is due to the fact that speaker generates signals in order to be understood/recognized, therefore, an analysis tailored to the way that the human ear works yields better performance.

**Prosodic:** Prosodic features are measures of stress, accent and intonation. The easiest way to estimate them is by means of pitch, energy, and duration information. Energy and pitch can be used in a similar way than the short-term characteristics of the previous level with a GMM model. Although these features by its own do not provide as good results as spectral features, some improvement can be achieved combining both kinds of features. Obviously different levels of data fusion can be used [6]. On the other hand, there are more potential using long-term characteristics. For instance, human beings trying to imitate the voice of another person usually try to replicate energy and pitch dynamics, rather than instantaneous values. Thus, it is clear that this approach has potential. Figure 4 shows an example of speech sentence and its intensity and pitch contours. This information has been extracted using the Praat software, which can be downloaded from [9]. The use of prosodic information can improve the robustness of the system, in the sense that it is less affected by the transmission channel than the spectral characteristics, and therefore it is a potential candidate feature to be used as a complement of the spectral information in applications were the microphone can change or the transmission channel is different from the one used in the training phase. The prosodic features can be used at two levels, in the lower one, one can use the direct values of the pitch, energy or duration, at a higher level, the system might compute co-occurrence probabilities of certain recurrent patterns and check them at the recognition phase.

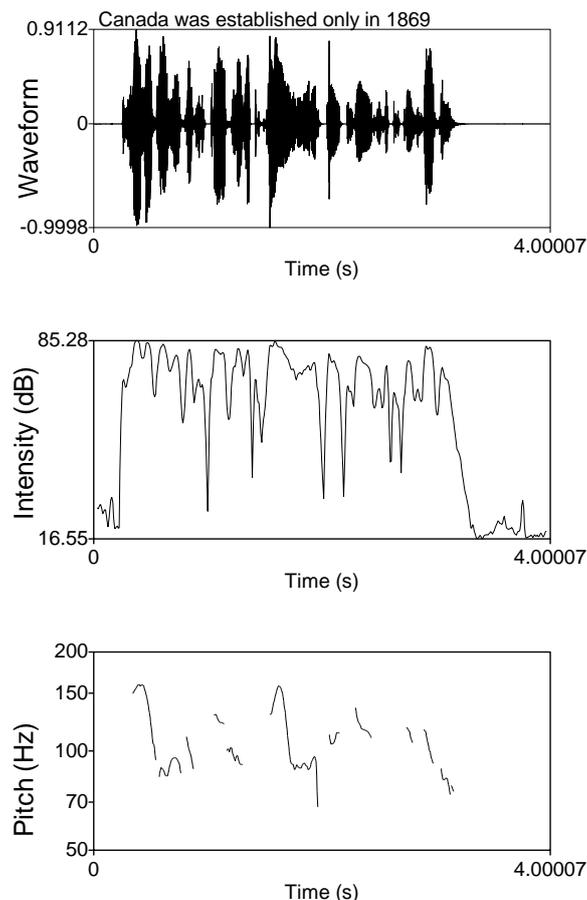

**Figure 4.** Speech sentence "Canada was established only in 1869" and its intensity and pitch contour. Although talking the same sentence, different speakers would produce different patterns, i.e. syllable duration, profile of the pitch curve.

**Phonetic:** it is possible to characterize speaker-specific pronunciations and speaking patterns using phone sequences. It is known that same phonemes can be pronounced in different ways without changing the semantics of an utterance. This variability in the pronunciation of a given phoneme can be used by recognizing each variant of each phoneme and afterwards comparing the frequency of co-occurrence of the phonemes of an utterance (N-grams of phone sequences), with the N-grams of each speaker. This might capture the dialectal characteristics of the speaker, which might include geographical and cultural traits. The models can consist of N-grams of phone sequences. A disadvantage of this method is the need of an automatic speech recognition system, and the need to model the confusion matrix (i.e. the probability that a given phoneme is confused by another one). In any case, as there are available dialectical databases [10] for the main languages, the use of this kind of information is nowadays feasible.

**Idiolectal (synthactical):** Recent work by G. Doddington [11] has found useful speaker information using sequences of recognized words. These sequences are called n-grams, and as was explained above, consists of the statistics of co-occurrence of n consecutive words. They reflect the way of using the language by a given speaker. The idea is to recognize speakers by their word usage. It is well known that some persons use and abuse of several words. Sometimes when we try to imitate them we do not need to emulate their sound neither their intonation. Just repeating their "favorite" words is enough. The algorithm consists of working out n-grams from speaker training and testing data. For recognition, a score is derived from both n-grams (using for instance the Viterbi algorithm). This kind of information is a step further than classical systems, because we add a new element to the classical security systems (something we have, we know or we are): something we do. A strong point of this method is that it does not only take into account the use of vocabulary specific to the user, but also the context, and short time dependence between words, which is more difficult to imitate.

**Dialogic:** When we have a dialog with two or more speakers, we would like to segmentate the parts that correspond to each speaker. Conversational patterns are useful for determining when speaker change has occurred in a speech signal (segmentation) and for grouping together speech segments from the same speaker (clustering).

The integration of different levels of information, such as the spectral, phonological, prosodic or syntactical is difficult due to the heterogeneity of the features. Different techniques are available for combining different information with the adequate weighting of the evidences and if possible the integration has to be robust with respect to the failure of one of the features. A common framework can be a bayesian modeling [12], but there are also other techniques such as data fusion, neural nets, etc.

## CONCLUSIONS

In this paper we have presented a review of the state of the art, and promising future lines on research and applications of speaker recognition. In the last years, improvements in the technology related to automatic speech recognition and the availability of a wide range of databases, has given the possibility of introducing high level features into the speaker recognition systems. Thus it is possible to use phonological aspects specific of the speaker or dialectical aspects which might model the region/background of the speaker as well as his/her educational background. Also the use of statistical grammar modelling can take into account the different word co-occurrence, of each speaker. An important aspect is the fact that these new possibilities for improving the speaker recognition systems have to be integrated in order to benefit of the higher levels of information that are available nowadays.

In [1] we stated different possibilities for fooling a biometric system. In addition, we presented some strategies to overcome these vulnerabilities. For speaker recognition, the main possibilities include:
a) A playback of a previous recording from the genuine speaker.
b) A hacker trying to imitate the genuine speech (may be with the help of a voice conversion system)
c) An artificial speech synthesizer trying to imitate the genuine speech

In this paper, we have summarized the state-of-the-art in speaker recognition, and we can support the following ideas:
a) A playback is worthless for text-dependent systems, where the speaker must pronounce a different sentence in each attempt to enter. For text-independent, a watermark (time stamp) can be incrusted inside the speech signal, in order to set up and expire-date.
b) The human voice is a complex mechanism, which can be imitated by experts. However, it is extremely difficult to replicate all the different levels of information contained in it. Thus, a data extraction at different levels and some kind of data fusion can make almost impossible to imitate a given speaker in all the levels. Hence, a proper designed recognizer can be safer that it seems.

## REFERENCES

[1] M. Faundez-Zanuy "On the vulnerability of biometric security systems" IEEE Aerospace and Electronic Systems Magazine Vol.19 nº 6, pp.3-8, June 2004
[2] M. Faundez-Zanuy "Biometric recognition: why not massively adopted yet?". IEEE Aerospace and Electronic Systems Magazine. 2004. In press.

[3] Martin A., Doddington G., Kamm T., Ordowski M., and Przybocki M., "The DET curve in assessment of detection performance", V. 4, pp.1895-1898, European speech Processing Conference Eurospeech 1997

[4] S. Furui *Digital Speech Processing, synthesis, and recognition*., Marcel Dekker, 1989.

[5] J. P. Campbell, D. A. Reynolds and R. B. Dunn "Fusing high- and low-level features for speaker recognition". Eurospeech 2003 Geneva.

[6] M. Faundez-Zanuy "Data fusion in biometrics". IEEE Aerospace and Electronic Systems Magazine. 2004. In press.

[7] D. A. Reynolds , R. C. Rose "Robust text-independent speaker identification using Gaussian mixture speaker models". IEEE Trans. On Speech and Audio Processing, Vol. 3 No 1, pp. 72-83 January 1995

[8] Cristianini, N., Shawe-Taylor, J., *An Introduction to Support Vector Machines*, Cambridge University Press, (2000).

[9] http://www.praat.org

[10] J. Ortega-García, J. González-Rodríguez and V. Marrero-Aguiar"AHUMADA: A Large Speech Corpus in Spanish for Speaker Characterization and Identification". Speech communication Vol. 31 (2000), pp. 255-264, June 2000

[11] G. Doddington, "Speaker Recognition based on Idiolectal Differences between Speakers," Eurospeech, vol. 4, p. 2521-2524, Aalborg 2001

[12] C. D. Manning, H. Schtze *Foundations of Statistical Natural Language Processing*, MIT Press; 1st edition (June 18, 1999).